  \documentstyle[prd, preprint,eqsecnum,aps]{revtex}  
  \begin{document}
  \draft
   \preprint{\vbox{
  \hbox{CTP-TAMU-14/99}
  \hbox{SINP-TNP/99-5}
  \hbox{hep-th/9904112}
  }}

  \title{$(m,\,n)$-String-Like Dp-Brane Bound States \\}

  \author {J. X. Lu$^1$ and Shibaji Roy$^2$}
  \address{$^1$Center for Theoretical Physics,
  Department of Physics\\
  Texas A\&M University, College Station, TX 77843\\
  E-mail: jxlu@rainbow.physics.tamu.edu\\
  and\\ $^2$Saha Institute of Nuclear Physics\\
  1/AF Bidhannagar, Calcutta 700 064, India\\E-mail: roy@tnp.saha.ernet.in}

  \date{\today}
  \maketitle
  
  \begin{abstract}
  
  An $(m,\,n)$-string bound state (with $m$, $n$ relatively prime integers)
  in type IIB string theory can be interpreted from the D-string worldsheet
  point of view as $n$ D-strings carrying $m$ units of quantized 
  electric flux or quantized electric field. We argue, from the D-brane 
  worldvolume  point of view, that similar Dp-brane bound states should 
  also exist for $ 2 \le p \le 8$ in both type IIA (when $p$ is even) and 
  type IIB (when $p$ is odd) string theories. As in $p = 1$ case, 
  these bound states can each be interpreted as $n$ Dp-branes carrying $m$ 
  units of quantized constant electric field.  In particular, they all 
  preserve one half of the spacetime supersymmetries.       
  
  \end{abstract}
  \newpage

  \section{Introduction\protect\\}
  \label{sec:intro}
  Polchinski's seminal work \cite{pol} on D-brane has dramatically changed 
our view on 
  perturbative 
  superstrings. Yet, we can use many tools developed in the perturbative 
  framework of superstrings to do D-brane calculations. These help us, at
  least in certain cases, to
  attack some hard problems in physics such as the information loss puzzle
and the entropy problem in black hole physics. The D-brane picture is also 
  the basis
  for the recent $AdS/CFT$ conjectures of Maldacena\cite{mal}. 
By definition, a D-brane is a
  hypersurface carrying a RR charge in type II string theory on which 
an open string can end.  
  From the D-brane worldvolume point of view, such an ending of a 
fundamental string (for
  short, F-string)  is characterized
  by the non-vanishing U(1) gauge field strength on the brane at least 
  in the low energy limit.  A configuration of an F-string ending 
on a Dp-brane 
  for every allowable $p$ can actually be BPS saturated, 
preserving a quarter of
  the spacetime supersymmetries. At the linearized approximation, 
this has been 
  demonstrated by Callan and Maldacena \cite{calm} for $p \ge 2$ cases and by 
  Dasgupta and Mukhi\cite{dasm} for $p = 1$ case. The 
interpretations for $p \ge 2$ and
  $p = 1$ cases are, however, quite different. In the former case, 
the excitation of a 
  worldvolume scalar field along a transverse direction is 
interpreted as the  
  F-string attached to the Dp-brane.  Whereas, for the latter one, 
the excitation of this scalar field due to the introduction of an
  F-string ending indicates that one half of the original D-string 
must bend rigidly to form
   a 3-string junction.  
   
  	In spite of our reasonably well understanding of an 
F-string ending on a Dp-brane from 
  the worldvolume point of view,  our understanding of this same ending 
from the spacetime point of view is still
  unsatisfactory\footnote{ For some very recent efforts in this
direction see \cite{you}.}. 
  The well-known $p$-brane solitonic solutions of type II supergravity
theories \cite{hows,dufl}, nowadays called Dp-branes,
  are merely hypersurfaces carrying RR charges
  each of which preserves one half of the spacetime supersymmetries of 
type II string theories. 
  The mass per unit $p$-brane volume for such a configuration carrying 
unit RR charge is just the 
  Dp-brane tension. This BPS configuration can also be described 
by its worldvolume Born-Infeld
  action in its simplest form with flat background and  vanishing 
worldvolume gauge 
  field strength which clearly indicates that 
  each of the spacetime Dp-brane solitonic solutions does not have 
an F-string
  ending on it (this also explains why they preserve 1/2 rather 
than 1/4 of the spacetime
  supersymmetries). 
  
  As just pointed out, a non-vanishing worldvolume gauge field 
strength is an indication of a string
  ending on the corresponding Dp-brane. In general we expect 
that such a configuration preserves
  a quarter of the spacetime supersymmetries. The question that we 
intend to address here is:  
  Does there exist a BPS state for each Dp-brane that has a
non-vanishing worldvolume gauge
  field strength and yet preserves one half of the spacetime 
supersymmetries? We will argue in this
  paper that the answer is yes based on known Dp-brane results 
and the 3-string junction. 
  Each of these BPS states is actually a non-threshold bound 
state of a Dp brane carrying certain units 
  of quantized constant electric field strength or a non-threshold (F, Dp) 
  bound state with F
  representing the F-strings. There actually exist more general non-threshold
  bound states. For example, by the type IIB S-duality, we should have
  D3 branes carrying both quantized constant electric and magnetic fields. 
  We will
  discuss the $p = 3$ case in this paper and others in the subsequent
  publications. In the following section, 
we will review relevant 
  Dp-brane results for the purpose of this paper. In section 3, 
we will present our arguments
  for the existence of such BPS states and conclude this paper.

  \section{Review of Some D-Brane Results\protect\\}
  \label{sec:RDP} 
  
  	This section is largely based on the discussion of BPS states of 
a fundamental 
  string (F-string) ending on a Dp-brane by Callan and 
Maldacena \cite{calm}
  for $p \ge 2$ and by Dasgupta and Mukhi \cite{dasm} for $p = 1$
    in the linearized approximation. The linear arguments should be 
trusted since we are here 
  interested only in BPS states.  As in \cite{calm}, we assume that 
the massless excitations of 
  a Dp-brane are described by the dimensional reduction of the 
10-dimensional supersymmetric 
  Maxwell theory. 
  The supersymmetry variation of the gaugino is
  \begin{equation}  
  \delta \chi = \Gamma^{MN} F_{MN} \epsilon,
  \label{eq:susy}
  \end{equation}
  where $M,N$ are the 10-dimensional indices. A BPS configuration is 
the one in which 
  $\delta \chi = 0$ for some non-vanishing killing spinor. 
The ending of an F-string
  on a Dp-brane is equivalent to placing a point charge on the brane. 
The Coulomb potential
  due to such a point charge will give rise to a non-vanishing $F_{0r}$ 
  with $r$ the radial coordinate of the $p$ spatial dimensions of 
the worldvolume. With a
  non-vanishing $F_{0r}$, it is obvious from Eq.\ (\ref{eq:susy}) 
and $\delta \chi = 0$ 
  that the existence of
  non-vanishing Killing spinors (i.e., the preservation of some unbroken 
supersymmetries), requires the excitation of
  one of the scalar fields, say $X^9$, such that 
$F_{9r} = - \partial_r X^9 = F_{0r}$. Then 
  $\delta \chi = 0$ can be expressed in a familiar form as 
  \begin{equation}
  (1 - \Gamma^0 \Gamma^9) \epsilon = 0,
  \label{eq:susyc}
  \end{equation}
  which says that one half of the worldvolume supersymmetries are 
broken by this
  configuration. In other words, this configuration of an F-string 
ending on a Dp-brane is 
  still a BPS state which preserves one half of the worldvolume 
supersymmetries or
  a quarter of the spacetime supersymmetries. It is easy to check that 
  $F_{0r} = F_{9r} = c_p (p -2)/r^{p -1}$ for $p > 2$, 
$F_{0r} = F_{9r} = c_2 / r$ for $p = 2$,
  and $F_{01} = F_{91} = c_1$ for $x^1 > 0$ and 
$F_{01} = F_{91} = 0$ for $x^1 < 0$\footnote{For
  concreteness, we assume that the original D-string is placed along 
the $x^1$-axis.}
   for $p =1$ satisfy the corresponding linearized equations of 
  motion, respectively. This has to be true to guarantee 
the existence of the corresponding BPS
  states. In the above, the constant $c_p$ is related to the 
point charge and can be fixed by
  some charge quantization which will be discussed later. 
  
To have a clear picture about an F-string ending on a Dp-brane, 
we need to examine 
  the above BPS configuration closely. The cases for $p \ge 2$ and $p = 1$ 
are quite different.
  So we discuss them separately. Let us discuss $p > 2$ first. 
Here we can solve $X^9$ from
  $F_{9r} = c_p (p - 2) / r^{p - 1} = - \partial_r X^9$ 
as $X^9 = c_p /r^{p - 2}$. As explained
  in \cite{calm}, the excitation of $X^9$ amounts to giving 
the brane a transverse `spike'
  protruding in the 9 direction and running off to infinity. 
This spike must be interpreted 
  as an F-string attached to the Dp-brane. Callan and Maldacena have 
shown that the energy
  change due to the introduction of a point charge to the 
Dp-brane worldvolume equals precisely
  to the F-string tension times $X^9$ which is the energy of an 
F-string if the spike is
  interpreted as the F-string. This also says that attaching an 
F-string to a Dp-brane 
  does not cost any energy which is not true in the case of $p = 1$.
  
  	The $p = 2$ case is not much different from $p > 2$ cases apart 
from the fact that $X^9$
  now behaves according to $X^9 = c_2~ {\rm ln}~ r/\delta$ 
with $\delta$ a small-distance cutoff
  rather than like a `spike'. This $X^9$, a sort of inverse `spike', 
should also be 
  interpreted as an F-string because of the underlying 
D-brane picture and the similar energy
  relation.\footnote{The energy change of D2 due to the 
introduction of a point charge to the worldvolume
  can also be expressed as the F-string tension times $X^9$. 
Here a large-distance cutoff needs to
  be introduced to make the calculation meaningful.}

  So far, we have considered only the single-center Coulomb 
solution in the
  above BPS states  describing an F-string ending on a Dp-brane 
for $p \ge
  2$. The BPS nature of these configurations allows multi-center
  solutions. For example, for $p > 2$, $X^9$ is now
  \begin{equation}
  X^9 = \sum_i \frac{c_p^i}{|\vec{r} - \vec{r}_i|^{p - 2}},
  \label{eq:mcs}
  \end{equation}
  where $c_p^i$ can be positive or negative, depending on to which 
side of
  the Dp-brane an F-string is attached. 
  This solution represents multiple strings along $X^9$ direction 
ending at
  arbitrary locations on the brane. This solution is still BPS, preserving
  also a quarter of the spacetime supersymmetries. The energy change of 
Dp-brane due
  to the endings of multiple strings is again equal to the summation
  of the F-string tension times individual F-string length and is
  independent of the locations of the end points. Therefore, no
  attachment energy is spent for such endings. These multi-center
  solutions are one of the important properties which we need in section 3.
  
  	The story for $p = 1$ case is quite different. As discussed in
  \cite{dasm}, the excitation of $X^9$ is no longer interpreted as an
  F-string ending on a D-string but as an indication that one
  side of the original
  infinitely long and straight D-string or (0,1)-string must be bent rigidly
  (with vanishing axion) with respect to the point on the
  D-string where a point charge is inserted. Let us look at it in some detail
  since the physics picture for this case consists of the starting point
  of our arguments for the existence of the Dp-brane bound states in the
  next section.
  In the presence of this point charge, Gauss' law in one spatial dimension
  states, in the case of vanishing axion, that $F_{01} = c_1$ for $x^1 > 0$
  and $F_{01} = 0$ for $x^1 < 0$\footnote{One can also have an 
alternative solution of
   $F_{01} = 0$ for $x^1 > 0$ and $F_{01} = - c_1$ for $x^1 < 0$.} 
when the original
  D-string or (0,1)-string is along $x^1$ axis. Unbroken susy condition 
  $F_{91} = - \partial_1 X^9 = F_{01}$ says
  \begin{eqnarray}
  X^9 &=& - c_1 x^1, \qquad x^1 > 0, \nonumber\\
      &=& 0,\qquad x^1 < 0.
  \label{eq:pqs}
  \end{eqnarray}
  
  Because of the special properties of 1 + 1 dimensional electrodynamics,
  the above solution is linearly increasing away from the inserted charge, 
in contrast
  to $p \ge 2$ cases. Before the work of Dasgupta and Mukhi, Aharony et al. 
  \cite{ahasy} concluded that three strings are allowed to meet at one point 
  provided there exist the corresponding couplings and the charges at the
  junction
  point are conserved. Schwarz \cite{sch} then went one step further to
  conjecture, based on his ($m$,$n$)-string \cite{schone} in type IIB theory, 
that
  there exists a BPS state for such 3-string junction provided the three
  strings are semi-infinite and the angles are chosen such that tensions,
  treated as vectors, add up to zero. With this, one should  not be surprised 
  about the above solution and the natural interpretation, as indicated
  already in \cite{sch} for an F-string ending on a D-string, that
  the insertion of the  point charge at the origin of the D-string 
causes one half
  of the string to bend rigidly. The solution itself does not spell out the
  ending of F-string or (1,0)-string. But a consistent picture requires that
  the point charge represents the ending  of a semi-infinitely long F-string 
or
  (1,0)-string (chosen here along positive $x^9$ direction) coming in
  perpendicular to the original D-string or
  (0,1)-string along $x^1$ direction. The bent segment described by
   Eq.\ (\ref{eq:pqs}) that goes out from the junction is a D-string 
carrying one 
   unit of quantized electric flux or Schwarz's (1,1)-string 
(or ($-1$,$-1$)-string
  depending on the orientation) in type IIB theory which follows 
from the charge conservation.
  The 3-string junction has also been studied by Sen\cite{sen} from
  spacetime point of view based on Schwarz's ($m$,$n$)-strings in 
type IIB theory. He
  showed that a 3-string junction indeed preserves 1/4 of the spacetime
  supersymmetries and a string network which also preserves 1/4 of the
  spacetime supersymmetries can actually be constructed using 3-string
  junctions as building blocks. Such a string network may, to our
  understanding, correspond to the multi-center solutions in $p \ge
  2$ cases. The energy change of the D-string due to the ending of an
  F-string, unlike the $p \ge 2$ cases, is no longer equal to the
  F-string tension times the attached F-string length, primarily due to
  the formation of the (1,1)-string bound state.
  
  	In summary, the 3-string junction, as the BPS state of an F-string
  ending on a D-string,  is just the consequence of
  D-brane picture, 1 + 1 dimensional electrodynamics and the
  non-perturbative SL(2,Z) strong-weak duality symmetry in 
  type IIB string theory.
  One important point to notice, which is well-known nowadays and will be
  useful in our later discussion, is that $m$ F-strings in the 
($m$,$n$)-string
  bound state in type IIB theory are just $m$ units of the quantized 
  electric flux. 
  
  	Therefore an ($m$,1)-string can be viewed as a D-string 
carrying $m$
  units of quantized electric flux. The ($m$,1)-string tension is
  $\sqrt{1/g^2 + m^2} T_f$ with $T_f = 1/{2\pi\alpha'}$ the F-string 
tension. For
  small string coupling $g$ and small $m$, the ($m$,1)-string tension can be
  approximated as $(1/g + g m^2/2) T_f$. Therefore, $(g m^2/2) T_f$ should 
  correspond to the linearized energy per unit length 
  of the worldsheet constant gauge field 
  strength $F_{01}$, i.e., $ ((2\pi \alpha'F_{01})^2 /2g)T_f$. So 
  we have $F_{01} = g m T_f $ which fixes $c_1 = g T_f$ for a single F-string. 
  By
  T-dualities, the electric field $F$ due to the ending of F-strings 
  on a Dp-brane is quantized according to
  \begin{equation}
  \frac{1}{(2\pi)^{p - 2} \alpha'^{(p - 3)/2}} \int_{S_{p - 1}} \ast F = g m,
  \label{eq:fq}
  \end{equation}
  where $\ast$ denotes the Hodge dual in the worldvolume. 
  This is precisely the condition used in \cite{calm} to fix the constant
  $c_p$.

  \section{${\rm D}_p$ Brane Bound States\protect\\}
  \label{sec:dpb}
  
  	Until now, we understand that the obvious reason for the existence
  of ($m$,$n$)-strings in non-perturbative type IIB string theory 
is the SL(2,Z)
  strong-weak duality symmetry under which the NSNS and RR 2-form potentials 
  transform as a doublet. However, as we will explain below, an 
($m$,$n$)-string 
  bound state is not special at all if it is  interpreted as $n$ D-strings
  carrying $m$ units of quantized electric flux or field strength 
  as discussed in the previous
  section. We will argue in this section that such a kind of bound states,
  i.e., a Dp-brane carrying certain units of quantized electric flux  or  field
  lines,
  actually exist for all Dp branes for $1 \le p \le 8$. All these bound states
  are BPS saturated and 
  preserve one half of the spacetime  supersymmetries just like an
  ($m$,$n$)-string. The fact that the ($m$,$n$)-strings were 
discovered earlier is
  because they can be easily recognized in the non-perturbative type IIB string 
  theory and  it happens that $m$ units of quantized electric flux or field
  strength can 
  be identified as $m$ F-strings.
  
  	Now to present our argument let us begin with a 3-string junction. 
Without loss of generality and
  for simplicity, let us focus on an F-string  ending on a D-string
   with zero axion. As discussed in the previous section, the third
  string is a D-string carrying one unit of quantized electric flux. 
  This is a stable BPS configuration which preserves a quarter of 
  the spacetime supersymmetries. 
  
  Suppose that we do not have an {\it a priori} knowledge of 
  Schwarz's ($m$,$n$)-strings and we do the linear study as Dasgupta and Mukhi 
  \cite{dasm} described in the previous section. The D-brane picture makes it
  certain that there must exist a stable BPS configuration of 
an F-string ending
  on a D-string which preserves  a quarter of the 
spacetime supersymmetries. So we
  must conclude from our linear analysis that this BPS state is a 3-string
  junction. The F-string remains as an F-string in the junction 
but the electric 
  charge at the end of the F-string will create a constant electric 
field or flux
  flowing along either side of the D-string with respect to the end point.
  At the final stable state, one side of the D-string remains as the original
  D-string but the other side becomes a D-string carrying one unit of quantized
  electric flux. This appears as 3 different kinds of strings 
meeting at one point.
  
  	We know that the D-string carrying one unit of quantized 
  electric flux or field strength in the 3-string junction is semi-infinite.
   Now let us push the 
  junction point to spatial infinity in such a way that the 
D-string carrying one
  unit of quantized electric flux is along one of the axes while 
the F-string and
the D-string are all at spatial infinity. To a local observer, 
  this D-string carrying one unit of quantized electric flux must appear to 
  be a stable BPS configuration\footnote{If one thinks carefully, 
  each of the  two ends of an ($m$,$n$)-string of Schwarz at 
spatial infinity must be either 
  associated with a 3-string junction or attached to any other 
allowable object.}.   
  Further, we must conclude that the D-string carrying certain units 
of quantized 
  electric flux must be a BPS one preserving one half of the spacetime 
  supersymmetries based on the facts that the 3-string junction preserves a 
  quarter of the spacetime supersymmetries and there exist BPS saturated 
  configurations for both the F-string and D-string each 
preserving one half of 
  the spacetime supersymmetries.
  In the 3-string junction, we must also conclude that supersymmetry conditions
  from any two constituent strings can be independent and the supersymmetry
  conditions from the remaining string must be related to those from the other 
  two strings\footnote{ We know that these are all true from Schwarz's 
($m$,$n$)-strings
  \cite{schone} and Sen's
  analysis of spacetime supersymmetry for 3-string junctions\cite{sen}.}.
  
  	So we conclude  that there exist a bound state of $n$ D-strings 
  carrying $m$ units of quantized electric flux based on 
D-brane picture, charge 
  conservation and the linear study discussed in the previous section. We now
  know that this bound state is just Schwarz's ($m$,$n$)-string
  which provides one way to identify one unit of quantized electric flux or
  field line  as
  an F-string\footnote{Another way of such an interpretation is given in
  \cite{polb}. }. 
   The only thing special for the bound state of 
$n$ D-strings carrying 
  $m$ units of quantized electric flux   is the  1 + 1
  dimensional electrodynamics which states that the gauge field strength is
  constant on one side of a point charge.  If we can
  consistently have a constant electric field 
  in a Dp-brane worldvolume, we find no reason that a Dp-brane 
carrying certain
  units of quantized electric flux should not exist from the above discussion.

  	To make our arguments for the existence of such Dp-brane bound
  states clear, let us first consider a specific $p = 3$ case. 
We take $p = 3$ partially
  because of the current fashion of $AdS_5/CFT_4$ correspondence and partially
  because of the familiarity of the 1 + 3 dimensional electrodynamics. We will
  discuss the general cases  for $ 1 \le p \le 8$ afterwards.
  
  In the case of D3-brane, we do not have the property of 1 + 1 dimensional
  electrodynamics. In general, when an F-string ends on a D3-brane, 
the F-string 
  will be spike-like, not rigid, near the end point. But this will not
  prevent us from doing the same as we did above for $p = 1$ case. 
As we will see, 
  insisting a constant electric field in any finite spatial region of 
  worldvolume in a consistent fashion will automatically push 
the endings of F-strings to spatial infinity.
  Therefore, the `spike' will appear to be a rigid F-string  to any finite 
region of space.
   
   The first
  question is
  what kind of electric charge distribution in 1 + 3 dimensions 
gives rise to a 
  constant electric field\footnote{It happens in this case that we can also
  have a bound state of a D3 brane carrying certain units of 
quantized constant magnetic field by the Type IIB S-duality. There actually 
exist such bound states for $2 \le p \le 8$ \cite{bremm}.  For p = 3, we can have 
a bound state of a D3 brane carrying both quantized constant electric and 
magnetic 
fields. We will discuss the p = 3 bound states later in this section. There 
actually exist similar and more general bound states which will be 
discussed in forthcoming papers\cite{lurthree,lurfour}.}.
  We know that a uniform 2-d surface charge distribution will do the job. 
The next
  question is where this surface should be placed. 
When we say a constant electric
  field, we mean that the field is 
  constant not only in magnitude but also in direction in any finite region of
  space. So we have to place this charge surface at spatial infinity. 
Otherwise,
  the direction of the electric field will be opposite on the two sides 
of the 
  surface. For concreteness, let us say that we 
label $x^1, x^2$ and $x^3$ as the 3-space
  of D3 brane and take the charge surface as $x^2 x^3$-plane and place it at
  $x^1 = - \infty$. Now where does the surface charge come from? 
It all comes from the 
  endings of parallel NSNS-strings, say along $x^9$ direction, 
on the $x^2x^3$-plane
  such that the resulting surface charge density is a constant. 
This is possible
  because of the existence of the multi-center solution discussed 
in the previous 
  section. Since these NSNS-strings are parallel to each other, 
the whole system
  is still a BPS one, preserving a quarter of the spacetime supersymmetries.
  Note that these endings of F-strings are now at spatial infinity.   
Therefore
  the `spikes' describing the endings of these F-strings have no influence on 
  the electric field
  in any finite region of space. So everything fits together nicely. 
In any finite
  region, we can detect only the D3-brane carrying a constant 
electric field in
  it. By the same token as in $p = 1$ case, we must conclude that 
this bound state  
  preserves also  one half of the spacetime supersymmetries. 
  
  Because the charge at the end of each of these NSNS-strings is quantized, we
  expect that the electric field should also be quantized. 
  If each of these NSNS-strings is $m$ F-strings, we should have 
here $F_{01} = g m T_f$
  with $g$ the corresponding string coupling constant. 
This can be obtained by 
  T-dualities from the $F_{01} = g m T_f$ in $p = 1$ case\footnote{To be more
  precise, we T-dualize the D3 brane Born-Infeld action with flat background 
  and non-vanishing constant worldvolume field $F_{01}$ along $x^2$ and $x^3$
  directions. We then end up with a D-string Born-Infeld action. 
Therefore we can
  read $F_{01} = g m T_f$. Noticing the relationship between the exact tension
  and linearized tension for $p = 1$ case, we must have the tension for the D3 
  brane bound state as given in Eq.\ (\ref{eq:btension}) since 
the two cases are
  related to each other by T-dualities.}.
  
  	The discussion for a general $p$ for $ 2 \le p \le 8$ is not 
much different
  from the $p = 3$ case. To be concrete, let us take the spatial 
dimensions of a Dp-brane
  along $x^1,  \cdots, x^p$. The $(p - 1)$ dimensional surface 
with uniform charge 
  distribution resulting from the endings of parallel 
NSNS-strings, say, along 
  $x^9$-direction is taken as a $(p - 1)$-plane along 
$x^2, \cdots, x^p$ directions and
  is placed at $x^1 = - \infty$. Then the electric field resulting 
from this charge surface
  will be constant and along $x^1$-direction in any finite region of space. 
It is also
  quantized as $F_{01} = g m T_f$ for an NSNS-string (to be thought of as 
$m$ F-strings). The rest will 
  be the same as in the case of $p = 3$. Since $F_{01} = g m T_f$, 
we can use the corresponding
  Dp-brane action to determine the corresponding tension 
$T_p (m,n)$ describing $n$ Dp-branes 
  carrying $m$ units of quantized constant electric field  which is 
  \begin{equation}
  T_p (m,n) = \frac{T^p_0}{g}  \sqrt{n^2 + g^2 m^2},
  \label{eq:btension}
  \end{equation}
  where $T^p_0 = 1/(2\pi)^p \alpha'^{(p + 1)/2}$. 
This expression clearly indicates that the configuration of 
  $n$ Dp-branes carrying $m$ units of quantized constant electric field 
with $m$ and $n$
  relatively prime integers is a non-threshold  bound state.
  So we conclude that $n$ Dp-branes carrying $m$ units of quantized constant
electric field consist of  a BPS non-threshold 
  bound state which preserves one half of the spacetime supersymmetries.
  
  	Since the quantized electric flux or field lines can be interpreted 
as F-strings, 
  these bound states should  be identified with the (F, Dp) bound states which 
are also related  to the  ($m$,$n$)-string or (F, D1) by T-dualities along the 
transverse directions. But here we must be
  careful about the notation `F' in (F, Dp). This F actually represents an 
  infinite number of parallel NS-strings along, say, $x^1$ direction, which 
  are distributed evenly over a $(p-1)$-dimensional plane 
perpendicular to the $x^1$-axis 
  (or the strings).
  As indicated above, each of these NS-strings is $m$ F-strings 
if $F_{01} = g m T_f$. The tension
  formula Eq.\ (\ref{eq:btension}) implies that we should have 
one NS-string (or $m$ F-strings)
  per $(2\pi)^{p-1} \alpha'^{(p - 1)/2}$ area over 
  the above $(p - 1)$-plane. 
  Since T-dualities preserve supersymmetries,  we can see in a 
different way that 
  these bound states preserve one half of the spacetime supersymmetries 
since the 
  original (F,D1) preserves one half of the spacetime supersymmetries.
  We will use this identification and perform T-dualities to construct 
explicitly the 
  spacetime configurations for these bound states in a forthcoming 
paper\cite{lurtwo}. We will show there that the tension formula 
Eq.\ (\ref{eq:btension}) holds and there are indeed $m$ F-strings per
$(2\pi)^{p -1} \alpha'^{(p -1)/2}$  area of ($p -1$)-dimensions. 
The spacetime configurations for (F, Dp) for $p = 3, 4, 6$ have been given in
\cite{rust,grelpt,cosp}, respectively.
  
  Once we have the above, it should not be difficult to have a non-threshold 
  bound state of $n$ D3 branes carrying $q$ units of quantized 
constant magnetic field with $n, q$
  relatively prime.
 All we need is to replace the F-strings in the above for $p = 3$ 
case by D-strings.
 If we also choose the quantized constant magnetic field along 
the $x^1$-axis, we must have
 $F_{23} = q T_f$ from the discussion in \cite{calm} about 
a D-string ending on a D3 
 brane. The corresponding tension is
 \begin{equation}
 T_3 (q, n) = \frac{T_0^3}{g} \sqrt{n^2 + q^2}.
 \label{eq:d1d3t}
 \end{equation}
 This tension formula implies that the linearized approximation 
on the D3 brane 
 worldvolume is good only if $n \gg q$. This bound state 
should correspond to the
 so-called (D1, D3) bound state. Again, we should have an 
infinite number of D-strings
 in this bound state and there should be $q$ D-strings per 
$(2\pi)^2 \alpha'$ area over 
 the $x^2x^3$-plane.
 
 Similarly, if we replace the F-strings or D-strings by ($m$,$q$)-strings 
in the
 above, we should end up with a non-threshold bound state of $n$ 
D3 branes carrying $m$ units of
 quantized electric flux lines and $q$ units of quantized magnetic 
flux lines with any two of
 the three integers relatively prime. The tension for this bound state is
 \begin{equation}
 T_3 (m,q, n) = \frac{T^3_0}{g} \sqrt{n^2 + q^2 + g^2 m^2}.
 \label{eq:d3dft}
 \end{equation}
 The linearized approximation on the worldvolume is good if either 
$n \gg q, n \gg m$ for
 fixed  and finite $g$ or $n \gg q$ for small $g$ and finite $m$. 
We denote this bound state as
 ((F, D1),D3). We should also have an infinite number of 
($m$,$q$)-strings in this bound state.
 We also have one ($m$,$q$)-string per $(2\pi)^2 \alpha'$ area 
over the $x^2x^3$-plane.
 
 	In \cite{lurthree}, we will construct explicit configuration
for ((F,D1),D3) bound state which gives the (D1, D3) bound state as a special
case. We will confirm all the above mentioned properties for them. The spacetime 
configurations for (Dp, D(p + 2)) for $0 \le p \le 4$ have been given in 
\cite{bremm,grelpt,cosp}. 

Note added: After the submission of this paper to hep-th, we were informed that the
existence of the bound states of Dp branes carrying constant electric fields was also
discussed in \cite{arfs} but in a completely different approach of the mixed boundary
conditions.

  \acknowledgments
  JXL acknowledges the support of NSF Grant PHY-9722090.


\begin{references}
  \bibitem{pol} J. Polchinski, Phys. Rev. Lett. {\bf 75} (1995) 4724.
  
  \bibitem{mal} J. Maldacena, Adv. Theor. Math. Phys. {\bf 2} (1998) 231.
  
  \bibitem{calm} C. Callan and J. Maldacena, Nucl. Phys. {\bf B513} (1998) 198.
  
  \bibitem{dasm} K. Dasgupta and S. Mukhi, Phys. Lett. {\bf B423} (1998) 261.
  
  \bibitem{you} D. Youm, ``Localized Intersecting BPS Branes'', hep-th/9902208.
  
  \bibitem{hows} G. Horowitz and A. Strominger, 
Nucl. Phys. {\bf B360} (1991) 197.
  
  \bibitem{dufl} M. J. Duff, R. Khuri and J. X. Lu, Phys. Rept. {\bf 259}
  (1995) 213.
  
  \bibitem{ahasy} O. Aharony, J. Sonnenschein and S. Yankielowicz, Nucl. Phys.
  {\bf B474} (1996) 309.
  
  \bibitem{sch} J. Schwarz, ``Lectures on superstring and 
M theory dualities'',
  Nucl. Phys. Proc. Suppl. {\bf B55} (1997) 1.
  
  \bibitem{schone} J. Schwarz, Phys. Lett. {\bf B360} (1995) 
13 (see hep-th/9508143
  for the most recent revision of this paper).
  
  \bibitem{sen} A. Sen, JHEP 9803: 005 (1998).

  \bibitem{polb} J. Polchinski, {\it String Theory}, Vol.II (Cambridge
University Press, 1998).

\bibitem{bremm} J. Breckenridge, G. Michaud and R. Myers, 
Phys. Rev. {\bf D55} (1997) 6438.
  
  \bibitem{lurthree} J. X. Lu and S. Roy,``((F,D1),D3) bound
   state and its T-dual daughters" hep-th/9905014.
   
  
  \bibitem{lurfour} J. X. Lu and S. Roy, ``(F,D5) bound state, SL(2,Z)
invariance and their descendants in type IIA/IIB string theory" hep-th/9905056.
  
  \bibitem{lurtwo} J. X. Lu and S. Roy, ``Non-threshold (F,Dp) bound states"
   hep-th/9904129.
   
 \bibitem{rust} J. G. Russo and A. A. Tseytlin, 
Nucl. Phys. {\bf B490} (1997) 121.

\bibitem{grelpt} M.B. Green, N. D. Lambert, G. Papadopoulos and  P.K. Townsend,
Phys. Lett. {\bf B384} (1996) 86.

\bibitem{cosp} M. Costa and G. Papadopoulos, Nucl. Phys. {\bf B510} (1998) 217.

\bibitem{arfs} H. Arfaei and M. M. Sheikh Jabbari, Nucl. Phys. 
{\bf B526} (1998) 278.  
  
  \end{references}
  \end{document}